# Equivalence relations between the Cortie and Zürich sunspot group morphological classifications

V.M.S. Carrasco[1,2], L. Lefèvre[3], J.M. Vaquero[4] and M.C. Gallego[2]

[1] Centro de Geofísica de Évora, Universidade de Évora, Évora, Portugal.
[2] Departamento de Física, Universidad de Extremadura, Badajoz, Spain.
[3] Royal Observatory of Belgium, Brussels, Belgium.
[4] Departamento de Física, Universidad de Extremadura, Mérida, Spain [e-mail: jvaquero@unex.es].

**Abstract:** Catalogues of sunspots have been available with useful information about sunspots or sunspot groups for approximately the last 150 years. However, the task of merging these catalogues is not simple. In this paper, a method is suggested of converting the types of sunspot groups that were proposed by Cortie (1901) into the well-known Zürich types of sunspot groups. To achieve this, the sunspot catalogue of the Valencia University Observatory (from 1920 to 1928) was used in addition to the descriptions proposed by Cortie. To assess the quality of this conversion scheme, the Zürich type was computed from the Valencia catalogue, and the resulting contribution of each group type was compared to what can be found in other catalogues. The results show that the proposed scheme works well within the errors that are found in the different catalogues.

## 1. Introduction

Sunspots are the product of intense magnetic fields occurring inside the Sun emanating toward the surface. They are of different sizes (measured typically in millionths of solar hemispheres: MSH) and lifetimes (hours to months). The number of sunspots present on the solar surface follows a periodic behaviour of about eleven years called the solar-activity cycle (Schwabe cycle) (Bray and Loughhead, 1964). Naked-eye observations of sunspots have been recorded sporadically over the last two millennia (Vaquero, Gallego and García, 2002; Vaquero and Vázquez, 2009). Since 1610, sunspots have been observed almost systematically with the use of telescopes (Vaquero, 2007). There are several indices that describe the behaviour of sunspots, for example, the International and Group Sunspot Numbers (Clette *et al*., 2014), that are based on counting sunspots with relatively small telescopes. The study and analysis of sunspot classification can improve the understanding of the solar cycle (e.g. Lefèvre and Clette, 2011; Kilcik *et al*., 2011) and therefore of space climate and its effects on the Earth's climate system. Detailed information about sunspots is contained in catalogues of sunspots that span approximately the last 150 years (Casas and Vaquero, 2014; Lefèvre and Clette, 2014). Often, however, these catalogues are based on disparate standards, and thus have different classification schemes. Therefore, a fundamental aspect to finding how to merge the catalogues is to find equivalence relationships between them (Lefèvre and Clette, 2014).

Cortie (1901) established a sunspot group classification according to the group's shape and its evolution. Later, the Zürich sunspot classification was developed by Waldmeier (1947), by modifying Cortie's (1901) previous scheme. The poor correlation of this classification with parameters of solar activity, such as solar flares, led to its modification to find a new classification that could improve flare predictions. Thus, the McIntosh classification (McIntosh, 1990) was established with three components, in which the nomenclature used for each sunspot group type is *Zpc,* where *Z* corresponds to "Modified Zürich Classification" (to make the classification change easy for observers), and the other two components, *p* and *c*, reflect the main sunspot characteristics: the type, size, and symmetry of the penumbra and umbra; and the degree of compactness of the group. In addition, there are other types of classification for sunspots, for example, the magnetic classification from Mt Wilson that describes the magnetic patterns (Hale and Nicholson, 1938).

The aim of this paper is to establish equivalence relationships between the sunspot group classifications of Cortie and Zürich. In this way, one will be able to extract more information about sunspot groups by extending in time the existing database of sunspot classification. In Sec. 2, we describe the Cortie and Zürich classifications. In Sec. 3, we present our proposal of equivalence relationships between the classifications. We give an example using the data from the Valencia University Observatory sunspot catalogue in Sec. 4. Finally, Sec. 5 presents the conclusions of the study.

## 2. Cortie and Zürich classifications

2.1. CORTIE CLASSIFICATION

2.1.1. Description of classes based on Cortie (1901)

Cortie (1901) established a classification of sunspot groups which was based on a study of about 3500 sunspot drawings made in the Stonyhurst College Observatory during the last twenty years of the nineteenth century. This classification corresponded to the typical shapes and patterns observed in the sunspot groups. The aim of the Cortie classification was to try to describe the different phases through which a group of spots goes during its evolution. This classification was used in various observatories. Two examples were the Valencia University Observatory (Carrasco *et al*., 2014) and the Madrid Astronomical Observatory (Aparicio *et al*., 2014) in the first decades of the twentieth century. Cortie defined the following sunspot group types (Fig. 1):

Type I: One spot or a group of small spots.
Type IIa: Group with a two-spot formation whose leading spot is the principal.
Type IIb: Group with a two-spot formation whose trailing spot is the principal.

Type IIc: Group with a two-spot formation with both spots being principal.
Type IIIa: Train of spots with well-defined principal spots.
Type IIIb: Train of penumbral spots with irregular umbra and without well-defined principal spots.
Type IVa: Single spot with regular contour.
Type IVb: Single spot with regular contour and small companions.
Type IVc: Single spot with irregular contour.
Type IVd: Single spot with irregular contour and train of small companions.
Type IVe: Single spot with irregular contour and small companions but not in a train.
Type V: Irregular group with large spots.

2.1.2. Description of classes based on parameters from the catalogue of the Astronomical Observatory of Valencia University (Valencia catalogue)

The Astronomical Observatory of Valencia University (Spain) was founded in 1909. It developed a program of monitoring solar activity. The astronomer Tomás Almer Arnau observed the Sun from 1920 to 1928. The equipment of this observatory included a 152-mm aperture Grubb refractor telescope and a photographic camera. The sunspot observations were performed using photography in which images of the Sun were taken with a diameter of 10 cm. A sunspot catalogue during the period 1920–1928, with a gap in the years 1921 and 1922, was published by the observatory (Observatorio de Valencia, 1928b). It has a temporal coverage of 74.0% during the period of study. Carrasco *et al*. (2014) recently provided a machine-readable version of this catalogue. For the sunspot groups, the Valencia Observatory adopted the classification proposed by Cortie (1901).

Because the description given above in Sec. 2.1.1 is rather vague, it is necessary to find other ways to describe the classes. To understand the objective criteria used to classify spots and groups into the different Cortie categories, we use three different group parameters available from the Valencia Catalogue (Carrasco *et al*., 2014): the maximum size of the spots in the group (the size of the largest spot), the longitudinal extent of the group and the number of spots in each group.

In Fig. 2 (upper left panel), one can see that Class I is made up essentially of spots that are below 100 MSH in size. One can assume that this was what the observers at the time considered to be "small spots". Figure 3 shows that they extend from 0 to 15 degrees. It is also evident that the vast majority of Class I groups have only one spot (i.e., groups of longitudinal extent 0 in Fig. 3) or two spots. Class II is made up of exactly two spots which are larger than those of Class I (the largest ones are mostly between 100 and 200 MSH), and of extent about the same as Class I (between 0 and 15 degrees). Class III has usually more than three spots, and extends beyond 15 degrees. The maximum size of its spots is in majority above 80 MSH (more than 60%). Classes IV and V both have the maximum sizes of their largest spots extend up to 1200 MSH and above and peak well above 200 MSH: Class IV has groups with one or two spots, Class V starts at three spots, and they both extend beyond 15 degrees. Note, however, that the peak extent for Class IV is below ten degrees, and has a large component at zero (one spot only), while Class V comprises a variety of longitudinal extents. Classes III and V are very similar in terms of extent and number of spots; they only differ in the size of their spots.

2.2. ZÜRICH CLASSIFICATION

The Zúrich classification (Fig. 4) is a well-known sunspot classification used by observers worldwide. It was developed by Waldmeier in 1938, based on the previous sunspot group classification defined by Cortie (1901). The same as its predecessor, it sought to describe the evolutionary sequence of sunspot groups. The nine types defined by Waldmeier are:

Type A: One or more spots without penumbra, and without bipolar configuration.
Type B: Group of spots without penumbra, with bipolar configuration.
Type C: Bipolar group which contains one spot with penumbra.
Type D: Bipolar group whose main spots have penumbra; length less than 10°.
Type E: Bipolar group whose main spots have penumbra; length between 10° and 15°.
Type F: Large bipolar group, length greater than 15°.
Type G: Bipolar group containing penumbras and no small spots between the main spots; length greater than 10°.
Type H: Single spot with penumbra and diameter larger than 2.5°.
Type J: Single spot with penumbra and diameter smaller than 2.5°.

## 3. Equivalence between the Cortie and Zürich classifications

In this section, we propose a set of conditions that sunspot groups from a database containing the Cortie classifications need to meet for them to be assigned to the Zürich categories. The proposed scheme is based on the definitions used by Cortie (1901) and Waldmeier (1938, 1947), and on the details obtained from the Valencia catalogue. It was impossible to find a unique relationship between the two classifications without additional information. It was necessary to also know the heliographic longitude of each sunspot in a group to calculate the length of the group and the area or the diameter of the single sunspots. We think that the most appropriate criteria for the equivalence between these two systems are the following rules (see Table 1):
- If a group has a Cortie Type I classification, it will be labeled as Zurich Type A or B if it has no penumbra (size criteria). In the case that the length of the group, defined by the difference between the maximum and minimum heliographic longitudes of the sunspots belonging to the group, is $< 3°$, the group is classified as A. Otherwise the type will be B. If any of the spots show a penumbra groups will be classified as C, D, or E. If the penumbra appears only on one side of the group then it will be a C group. If it appears on both sides and the length is $< 10°$ then it will be a D group. Otherwise it will be an E group. For individual spots, if a group is Cortie Type I and its area is less than 50 MSH then it will be an A group, otherwise it will be a J group.

- The Cortie classification of a Type II group can only be of Zürich Types C, D, or G. For all Cortie subtypes (IIa, IIb, and IIc), if the length of the group is greater than 10° then it will be Zürich Type G as it contains exactly two spots; if the group has a longitudinal extension of less than 10° with penumbral spots on both sides then it will be D, and if the penumbra is on one side only then it will be C.
- The equivalence for any group with Cortie Type III (IIIa or IIIb) will be Zürich Type D (if length of the group is less than 10°), E (if the length of the group is between 10° and 15°), or F (if length of the group is greater or equal to 15°).
- For Cortie Type IV, there are many possible classes. Unipolar groups of sunspots with diameters less than 2.5° or group areas less than 250 MSH belonging to any subtype of Cortie Type IV (IVa, IVb, IVc, IVd, and IVe) will be classified as Zürich Type J. The classification will be H when the sunspots of this Cortie type are greater than or equal to those diameters or areas. For the IVb, IVd and IVe subtypes: sunspots with penumbra on one side of the group only will be classified as C; with penumbra on both sides, no spots in between, and a length greater than 10° will be G; otherwise, it will be Zürich Type D (if length of the group is less than 10°), E (if the length of the group is between 10° and 15°), or F (if length of the group is greater or equal to 15°).
- Groups of Cortie Type V will be assigned Zürich Type D (if length of the group is less than 10°), E (if the length of the group is between 10° and 15°), or F (if length of the group is greater or equal to 15°).

## 4. Case Study: The Valencia Sunspot Catalogue

Applying the conditions described above, one can obtain an equivalence between the Cortie and Zürich classifications. For the specific case of the Valencia Sunspot Catalogue, Fig. 5 shows the percentage of each type for both classifications of the records in the catalogue of the Valencia Observatory. The "no central spots" criterion for equivalence between Cortie Type IV and Zurich G means that no sunspot should be in the range $X \pm 0.2L$, where $X$ is the heliographic longitude of the centre of the group, and $L$ is the longitudinal extension of the group, defined by the difference between the maximum and minimum heliographic longitudes of the sunspots. Since it is not specified in the catalogue if the spots have penumbra or not, we considered that the criterion "penumbra on one side only" is met if at least one sunspot on one side surpasses 50 MSH and no sunspot on the other side reaches this value. This value is chosen following the work of Hathaway (2013) showing that the penumbra starts to develop at around 50 MSH for single spots. Below that limit, we shall consider it to be a simple spot without penumbra. In Cortie Type IV, we also distinguish unipolar groups when they have an extent less than 4° (mean value of the extent of this Cortie type). The greatest percentage of sunspots in the Zürich classification corresponds to Type J, which has its equivalents in several Cortie types. The lowest percentages correspond to Cortie Type IVe and Zurich G. Spotless days (SD) and days without a type being determined (WT) represent 4.7% and 0.2% of all observations, respectively. The great percentage of cases corresponding to Type I in the Cortie classification is more evenly distributed in the Zurich classification. Ananthakrishnan (1952) carried out a statistical study about the sunspot group types, which follow the morphological classification of Cortie (1901), recorded in Kodaikanal Observatory for the period 1903–1950. A comparison between the percentages for the different Cortie types recorded in Valencia and Kodaikanal Observatory for the coincident period 1920–1928 is shown in Table 2. In both observatories, the highest percentages are obtained for Type I and IV. The discrepancy between observatories lies in the types with low percentages. While for the Valencia Observatory the lowest percentage is recorded for Type III, in Kodaikanal Observatory it is for Type II. According to the types with low percentages (Type II, III and V), we can see that: i) in the Valencia catalogue, the percentage for Type II and III is similar and the Type V proportion significantly higher and ii) in Kodaikanal Observatory Type III and V show in similar proportions while Type II is significantly lower. This difference is mainly due to the observer's bias that is different at both locations. The current study, by taking into account the actual parameters of the groups, goes farther than Ananthakrishnan (1952) and thus obtains accurate, if empirical, relations between the Cortie and Zürich classification schemes.

Some of the values of the distribution of the Valencia catalogue groups in the Zürich classification match reasonably well with modern catalogue values (USET catalogue, Lefèvre, 2015, personal communication) or earlier work (Kleczek, 1953), as one observes in Table 3. There is a slight deficit in the percentage of B groups and a slight excess of the D, F, and J groups with respect to the other catalogues. Figure 6 shows the distribution of the average area of the groups in the resulting Zürich classification. It is very similar to what can be found in such catalogues as those of the RGO or USAF (Lefèvre and Clette, 2011; Kilcik *et al.*, 2011; Kilcik *et al.*, 2014).

Figure 7 shows the annual averages of the group number by Zürich classification types deduced for the catalogue of the Valencia Observatory. We have also plotted the Sunspot Number Index calculated from the Valencia Observatory data as: *VISN = 10 G + S*; where *G* and *S* are the number of observed groups and sunspots, respectively. This index takes its maximum value in 1928, and one can appreciate a Gnevyshev gap in that the value of *VISN* for 1926 is greater than that for 1927. The Types E (maximum annual value 0.31) and F (0.24) show the same pattern of behaviour. Types A (0.76), B (0.31), and C (0.49) have their maxima in 1926, as does the Valencia Sunspot Area Index (Carrasco *et al.*, 2014). The maximum for Type G (0.10) occurs in 1927. Types D (0.79), H (0.34), and J (0.90) present maxima in 1928, as does the International Sunspot Number Index.

## 5. Conclusions

More precise information about past solar activity provides us with a better understanding of our star, and a better ability to predict its behaviour in the future. In this paper, we have presented a method to convert the Cortie type, which appears in several historical catalogues of sunspots, to the Zürich type, which is more common in the catalogues of the second half of the twentieth century and has a precise scheme of conversion to the modern McIntosh classification. A conversion such as that described here is a necessary step for the homogenization and merging of different sunspot catalogues. We presented an example of this conversion using solar observations made at the Observatory of Valencia University. One of the drawbacks of the Cortie classification is the great abundance of Type I groups. This is solved by the conversion to the Zürich classification since these groups are then redistributed into several types. According to the proportion of the different Cortie classifications, the Valencia catalogue has a similar behaviour than that of the Kodaikanal catalogue for the coincident record period (1920-1928). We found small deficits or excesses in the percentages of some group Zürich classes in the Valencia catalogue, but do not deem them to be significant at this point since the Valencia catalogue spans only a few years, and variations in the different phases of the cycles can explain the level of discrepancy found. Furthermore, we have shown that the annual averages for each individual sunspot type according to the Zürich classification as recorded in the Valencia Astronomical Observatory have a similar behaviour from that of the solar cycle, with maxima between 1926 and 1928.

**Acknowledgements**

Support from the Junta de Extremadura (Research Group Grant No. GR10131), from the Ministerio de Economía y Competitividad of the Spanish Government (AYA2011-25945), and from the COST Action ES1005 TOSCA (www.tosca-cost.eu) is gratefully acknowledged. L.L. acknowledges funding from the European Community's Seventh Framework Programme (FP7-SPACE-2012-2) under Grant Agreement 313188 (SOLID project, projects.pmodwrc.ch/solid/).

**Compliance with Ethical Standards**

Conflict of Interest: The authors declare that they have no conflict of interest.

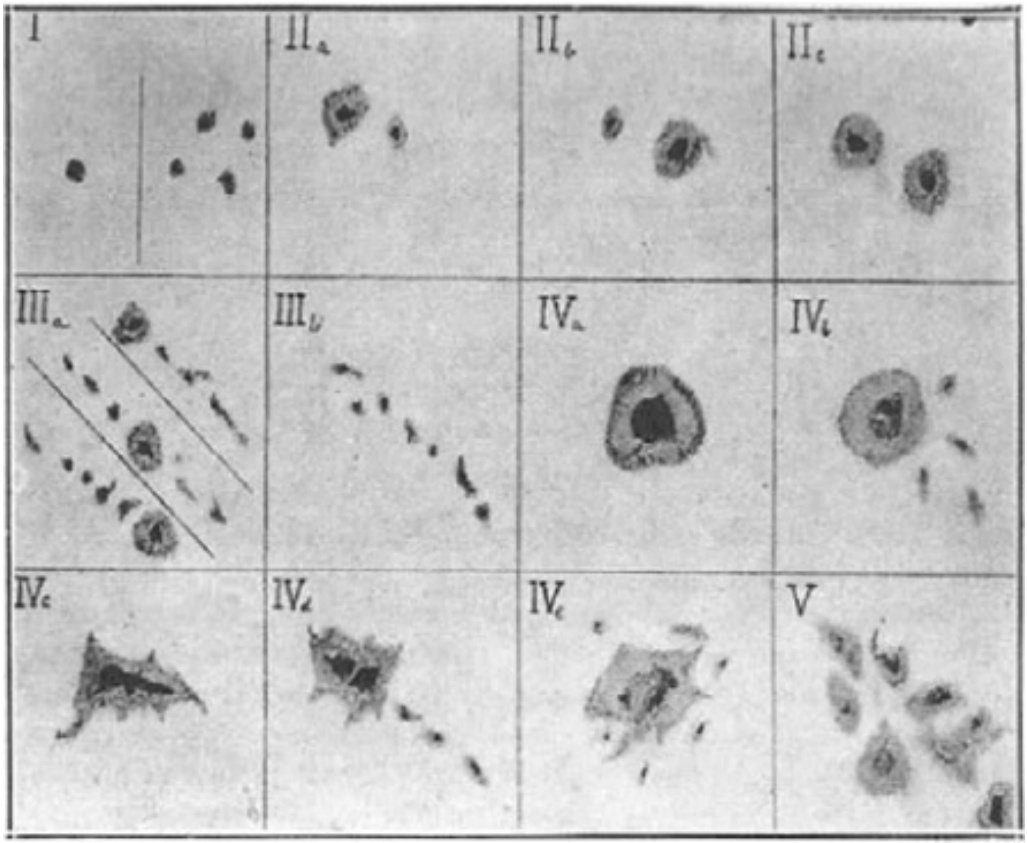

**Figure 1.** Examples of sunspot groups in the Cortie classification [Source: Observatorio de Valencia, 1928a].

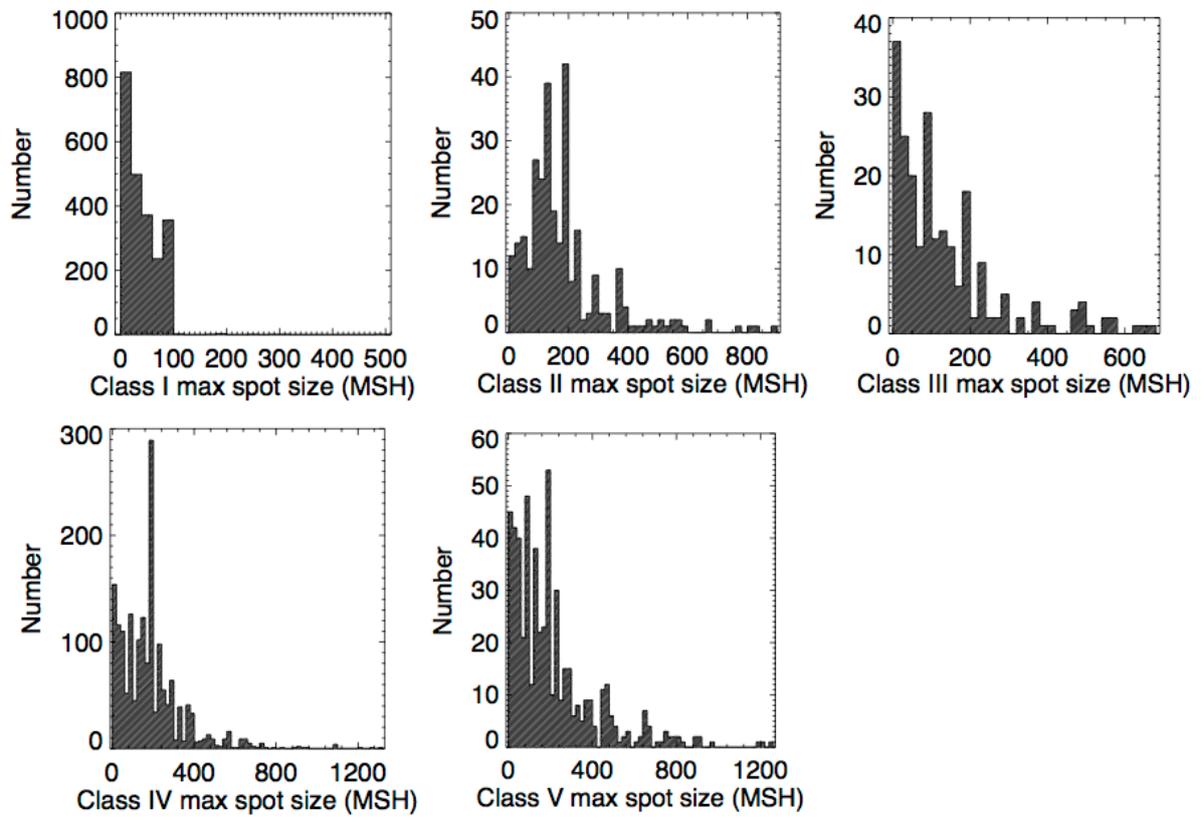

**Figure 2.** Size of the largest spot in the groups from the Valencia catalogue in Classes I, II, III, IV, and V (defined in the text).

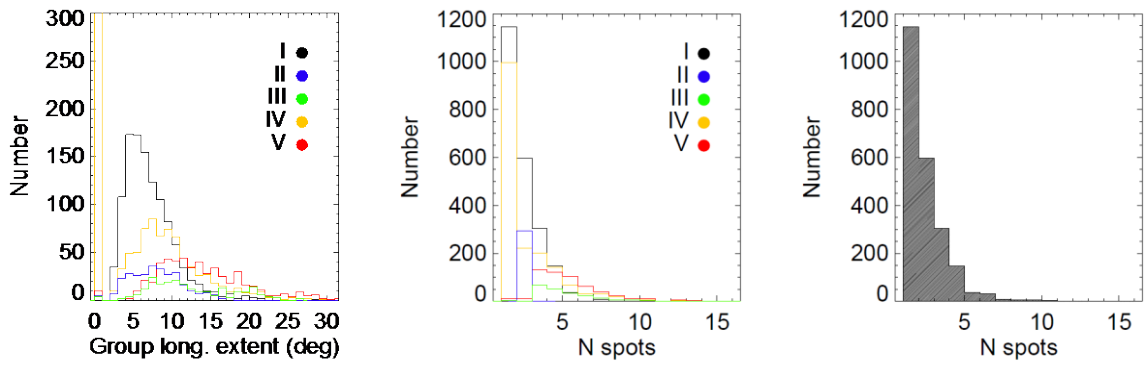

**Figure 3.** Left panel: Longitudinal extension of groups from the Valencia catalogue in the different classes defined in the text. Middle panel: Number of spots in groups from the Valencia catalogue in the different classes. Right panel: Number of spots in the groups of Class I.

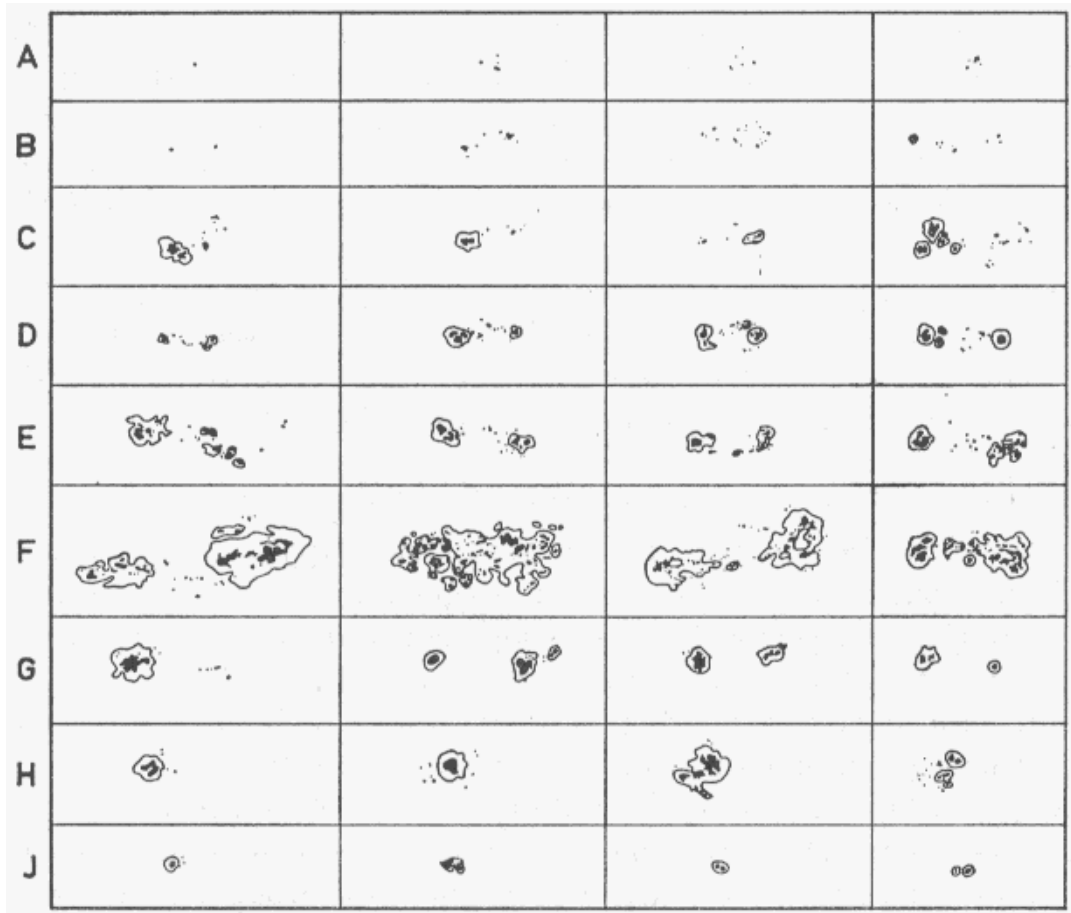

**Figure 4.** Examples of the Zurich classification of sunspot groups [source: Bray and Loughhead, 1964].

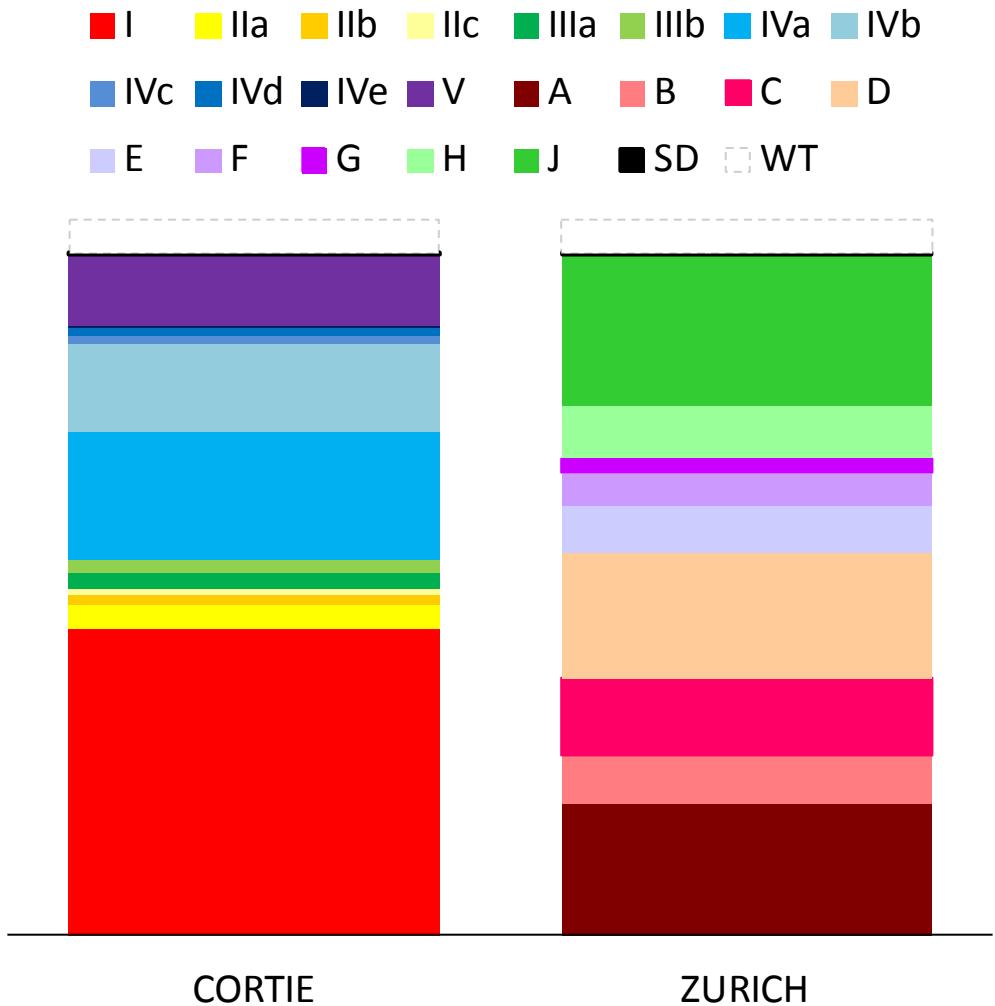

**Figure 5.** Distribution of the sunspot groups recorded in the Observatory of Valencia University according to the Cortie and Zürich classifications. SD (4.7%) and WT (0.2%) represent spotless days and group cases without classification, respectively. The percentages for the different types are: I (42.8%), IIa (3.3%), IIb (1.4%), IIc (0.8%), IIIa (2.2%), IIIb (2.0%), IVa (17.9%), IVb (12.1%), IVc (1.1%), IVd (1.1%), IVe (0.4%), V (10.0%) for Cortie class and A (18.4%), B (6.8%), C (10.7%), D (17.3%), E (6.9%), F (4.4%), G (2.1%), H (7.3%), J (21.2%) for Zürich class.

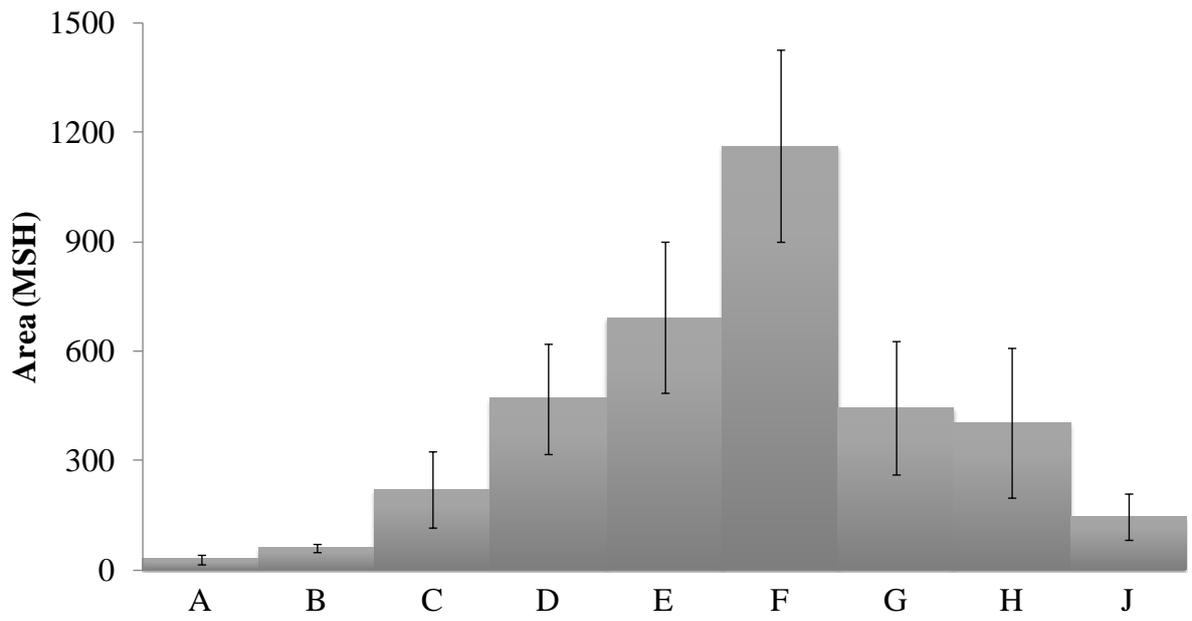

**Figure 6.** Average area of the groups in the Zürich classification deduced for the catalogue of the Valencia Observatory.

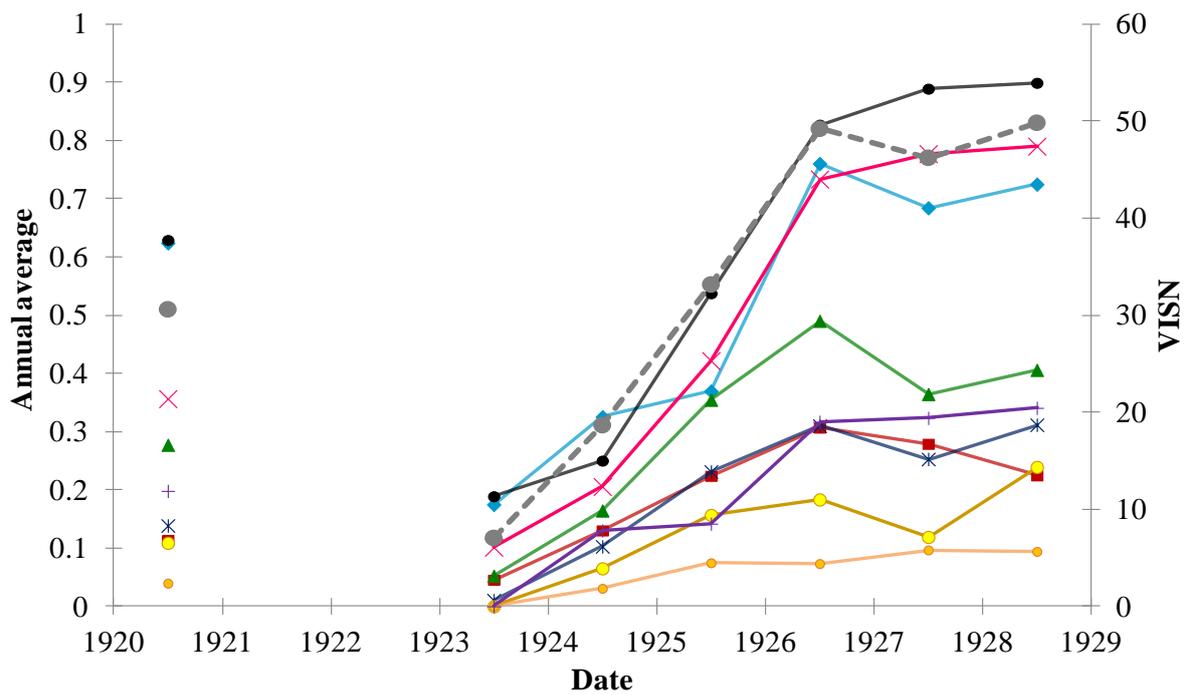

**Figure 7.** Annual average for A (blue), B (red), C (green), D (pink), E (dark blue), F (yellow), G (orange), H (purple) and J (black) types in the Valencia sunspot catalogue after applying the equivalence relations. Moreover, the sunspot number computed in this Observatory (VISN index, grey dashed line) is also represented.

**Table 1.** Summary of the equivalence criteria between the Cortie and Zürich classifications of sunspot groups. L is the extension of the group, D is the diameter of the Cortie Type IV sunspot, and A is the area of the group in units of millionths of the solar hemisphere (MSH).

| CORTIE | CRITERIA | ZURICH |
|---|---|---|
| I | No penumbra + L < 3° | A |
| I | No penumbra + L ≥ 3° | B |
| I | Penumbra on one side only | C |
| I | Penumbra on both sides + L < 10° | D |
| I | Penumbra on both sides + L ≥ 10° | E |
| I | Single spot + A < 50 MSH | A |
| I | Single spot + A ≥ 50 MSH | J |
| IIa,b | L ≥ 10° | G |
| IIa,b | Penumbra on one side only + L < 10° | C |
| IIa,b | Penumbra on both side + L < 10° | D |
| IIc | L ≥ 10° | G |
| IIc | L < 10° | D |
| III | L < 10° | D |
| III | 10° < L < 15° | E |
| III | L ≥ 15° | F |
| IVa,c | D < 2.5° or A < 250 MSH | J |
| IVa,c | D ≥ 2.5° or A ≥ 250 MSH | H |
| IVb,d,e | Penumbra on one side only + L ≥ 4° | C |
| IVb,d,e | Penumbra on both side + 4° < L < 10° | D |
| IVb,d,e | Penumbra on both side + 10° < L < 15° | E |
| IVb,d,e | Penumbra on both side + L ≥ 15° | F |
| IVb,d,e | No central spots + L ≥ 10° | G |
| IVb,d,e | L < 4° + D < 2.5° or A < 250 MSH | J |
| IVb,d,e | L < 4° + D ≥ 2.5° or A ≥ 250 MSH | H |
| V | L < 10° | D |
| V | 10° < L < 15° | E |
| V | L ≥ 15° | F |

**Table 2.** Percentages of the Cortie types in Valencia (this study) and Kodaikanal (Ananthakrishnan, 1952) Observatory for the coincident record period (1920-1928).

| CORTIE TYPE | VALENCIA | KODAIKANAL |
|---|---|---|
| I | 45.0 | 64.1 |
| II | 5.8 | 0.8 |
| III | 4.4 | 11.1 |
| IV | 34.2 | 22.5 |
| V | 10.6 | 1.5 |

**Table 3.** Percentages of the Zurich types in three different catalogues: USET (Lefèvre, 2015, personal communication), Kleczek (1953), and Valencia (this study).

| ZURICH TYPE | USET (1940-2014) | KLECZEK (1938-1950) | VALENCIA (1920-1928) |
|---|---|---|---|
| A | 18.9 | 28.7 | 19.3 |
| B | 14.4 | 11.2 | 7.2 |
| C | 17.8 | 12.3 | 11.3 |
| D | 15.6 | 11.1 | 18.1 |
| E | 4.8 | 7.6 | 7.3 |
| F | 1.3 | 2.0 | 4.6 |
| G | 2.7 | 5.1 | 2.2 |
| H | 5.3 | 8.0 | 7.7 |
| J | 19.2 | 13.8 | 22.3 |